\documentclass[12pt]{article}
\usepackage{amssymb}
\usepackage{amsmath}
\usepackage{amsfonts}

\newcommand{\be}{\begin{equation}}

\newcommand{\ee}{\end{equation}}
\newcommand{\bea}{\begin{eqnarray}}
\newcommand{\eea}{\end{eqnarray}}
\newcommand \tr {\hbox{Tr}}

\begin{document}

\title{Complementarity and additivity for depolarizing channels}
\author{Nilanjana Datta \\
Statistical Laboratory, Centre for Mathematical Sciences, \\
University of Cambridge\\
Alexander S. Holevo \\
Steklov Mathematical Institute}
\date{}
\maketitle

\begin{abstract}
In this paper we use the method of the paper \cite{holcc} to compute
complementary channels for certain important cases, such as depolarizing and
transpose-depolarizing channels. This method allows us to easily obtain the
minimal Kraus representations from non--minimal ones. We also study the
properties of the output purity of the tensor product of a channel and its
complement.
\end{abstract}

\section{\protect\bigskip Introduction}

In the recent paper \cite{holcc} complementarity between output and
environment of a quantum channel (or, more generally, CP map) was explored
in detail. It was observed that the output purity characteristics for
mutually complementary CP maps coincide, making the validity of the
mutiplicativity/additivity conjecture for a class of CP maps equivalent to
its validity for complementary maps. A similar observation was independently
made in \cite{rus} in the context of channels. In \cite{holcc} a regular
method for computation of complementary maps was proposed, thus providing an
efficient construction of new cases for the solution of the
multiplicativity/additivity problem. In this paper we use this method to
compute complementary channels for certain important cases, such as
depolarizing and transpose-depolarizing channels. This method easily yields
minimal Kraus representations from non--minimal ones. We also study the
properties of the output purity of the tensor product of a channel and its
complement.

Let us fix some notation. $\mathfrak{M}\left( \mathcal{H}\right) $ will denote the
algebra of all operators, and $\mathfrak{S}(\mathcal{H})$ -- the convex set of all
density operators (quantum states) in a finite-dimensional Hilbert space
$\mathcal{H} $. The output purity of a CP map $\Phi :\mathfrak{ M}\left(
\mathcal{H}  \right) \rightarrow \mathfrak{M}\left( \mathcal{H} ^{\prime }\right)
,$ is measured by the quantity
\begin{equation}
\nu _{p}(\Phi ):=\max_{\rho \in \mathfrak{S}(\mathcal{H})}\left\{ ||\Phi
(\rho )||_{p}\right\} ,\quad 1\leq p<\infty ,  \label{nup}
\end{equation}
where ${\displaystyle{||\Phi (\rho )||_{p}=\Bigl[\tr\left( \Phi (\rho
)\right) ^{p}\Bigr]^{1/p}}}$ is the $p$-norm of $\Phi (\rho )$, or
equivalently, by the minimal output R\'{e}nyi $p$-entropy
\begin{equation*}
\check{R}_{p}(\Phi )=-\frac{p}{p-1}\,\log \nu _{p}(\Phi ).
\end{equation*}
Recall that the R\'{e}nyi $p$-entropy of a density matrix $\sigma $, $p>1$,
is defined as
\begin{equation}
R_{p}(\sigma ):=-\frac{1}{p-1}\log \left( \tr\,\sigma ^{p}\right) =-\frac{p}{
p-1}\log ||\sigma ||_{p}.  \label{rendef}
\end{equation}
The R\'enyi entropies have the monotonicity property \cite{cup}
\begin{equation*}
R_{q}(\sigma )\leq R_{p}(\sigma ),\quad 1<p\leq q.
\end{equation*}
In the limit $p\rightarrow 1$, they converge monotonically and hence uniformly to
the von Neumann entropy $H(\sigma )=-\tr\,\sigma \log \sigma $. Therefore we can
extend the notation of the R\'{e}nyi entropy by letting $ R_{1}(\sigma ):=H(\sigma
)$. The minimal output R\'{e}nyi $p$-entropy of a channel $\Phi $ is then
\begin{equation*}
\check{R}_{p}(\Phi )=\min_{\rho \in \mathfrak{S}(\mathcal{H})}R_{p}(\Phi
(\rho )),
\end{equation*}
and for $p=1$ it is equal to the minimum output entropy
\begin{equation}
\check{H}(\Phi ):=\min_{\rho }\,H(\Phi (\rho )).  \label{minent1}
\end{equation}

\section{\protect\bigskip Representations of CP maps}

Given three Hilbert spaces $\mathcal{H}_{A},\mathcal{H}_{B},\mathcal{\ H}
_{C} $ and a linear operator $V:\mathcal{H}_{A}\rightarrow \mathcal{H}
_{B}\otimes \mathcal{H}_{C}$, the relations
\begin{equation}
\Phi (\rho )=\mathrm{Tr}_{\mathcal{H}_{C}}V\rho V^{\ast },\quad \tilde{\Phi}
(\rho )=\mathrm{Tr}_{\mathcal{H}_{B}}V\rho V^{\ast };\quad \rho \in
\mathfrak{M}\left( \mathcal{H}_{A}\right) ,  \label{compl}
\end{equation}
define two CP maps $\Phi :\mathfrak{M}\left( \mathcal{H}_{A}\right) \rightarrow
\mathfrak{M}\left( \mathcal{H}_{B}\right) ,$ $\tilde{\Phi}: \mathfrak{M}\left(
\mathcal{H}_{A}\right) \rightarrow \mathfrak{M}\left( \mathcal{H}_{C}\right) ,$
which will be called mutually \textit{complementary }. If $V$ is an isometry, both
maps are trace preserving (TP) i.e. channels.

For any linear map $\Phi :\mathfrak{M}\left( \mathcal{H}\right) \rightarrow
\mathfrak{M}\left( \mathcal{H}^{\prime }\right) $ the dual map $\Phi ^{\ast
}:\mathfrak{M}\left( \mathcal{H}^{\prime }\right) \rightarrow \mathfrak{M}
\left( \mathcal{H}\right) $ is defined by the formula
\begin{equation*}
\mathrm{Tr}\Phi (\rho )X=\mathrm{Tr}\rho \Phi ^{\ast }(X);\quad \rho \in
\mathfrak{M}\left( \mathcal{H}\right) ,X\in \mathfrak{M}\left( \mathcal{H}
^{\prime }\right) .
\end{equation*}
If $\Phi $ is CP, then $\Phi ^{\ast }$ is also CP. Relations (\ref{compl})
are equivalent to
\begin{equation}
\Phi ^{\ast }(X)=V^{\ast }(X\otimes I_{C})V;\quad X\in \mathfrak{M}\left(
\mathcal{H}_{B}\right) ,
\end{equation}
\begin{equation}  \label{df}
\tilde{\Phi}^{\ast }(X)=V^{\ast }(I_{B}\otimes X)V;\quad X\in \mathfrak{M} \left(
\mathcal{H}_{C}\right),
\end{equation}
where $I$ is the identity operator in the corresponding Hilbert space.
Considering $\tilde{\Phi}$ as dual to CP map $\tilde{\Phi}^{\ast },$ we
conclude that there should also be a representation of the form
\begin{equation}
\tilde{\Phi}(\rho )=S_{C}(\rho \otimes I_{B})S_{C}^{\ast },  \label{durep}
\end{equation}
where $S_{C}:{\mathcal{H}}_{A}\otimes {\mathcal{H}}_{B}\rightarrow
{\mathcal{H}}_{C}$ (in the case of channel $\mathrm{Tr}_{{\mathcal{H}}
_{B}}S_{C}^{\ast }S_{C}=I_{A}).$ There is a simple general relation between this
representation and the second formula in (\ref{compl}) for an arbitrary CP map;
namely, given $V:{\mathcal{H}}_{A}\rightarrow {\mathcal{H}} _{B}\otimes
{\mathcal{H}}_{C}$ choose an orthonormal basis $\left\{ e_{j}^{B}\right\} $ in
${\mathcal{H}}_{B}$ and define $S_C:{\mathcal{H}} _{A}\otimes {\mathcal{H}}
_{B}\rightarrow {\mathcal{H}}_{C}$ by the relation $\langle
e_{j}^{B}|V=S_{C}|e_{j}^{B}\rangle ,$ or, more precisely,
\begin{equation*}
\langle \bar{\psi}_{B}\otimes \psi _{C}|V|\psi _{A}\rangle =\langle \psi
_{C}|S_{C}|\psi _{A}\otimes \psi _{B}\rangle ,
\end{equation*}
where $\bar{\psi}_{B}$ denotes the complex conjugate of ${\psi}_B$ in the
basis $\left\{ e_{j}^{B}\right\} .$ Alternatively, introducing the maximally
entangled vector
\begin{equation*}
|\Omega ^{BB}\rangle =\frac{1}{\sqrt{d_{B}}}\sum_{j=1}^{d_{B}}|e_{j}^{B}
\rangle \otimes |e_{j}^{B}\rangle
\end{equation*}
in ${\mathcal{H}}_{B}\otimes {\mathcal{H}}_{B},$ we have the reciprocity
relations
\begin{equation*}
S_{C}=\sqrt{d_{B}}\langle \Omega ^{BB}|(V\otimes I_{B});\quad V=\sqrt{d_{B}}
(I_{B}\otimes S_{C})|\Omega ^{BB}\rangle .
\end{equation*}

The representation (\ref{durep}) is in fact nothing but the dual form (\ref {df})
of the Stinespring representation for the map $\tilde{\Phi}$, if it is considered
(somewhat \textquotedblleft unnaturally\textquotedblright ) as a map in the
Heisenberg-- rather than in the Schr\"{o}dinger picture. To give a kind of physical
interpretation to the representation (\ref{durep}), consider the polar
decomposition $S_{C}=|S_{C}^{\ast }|W,$ where $ |S_{C}^{\ast
}|=\sqrt{S_{C}S_{C}^{\ast }}$ is a Hermitian operator in ${ \mathcal{H}}_{C}$ and
$W:{\mathcal{H}}_{A}\otimes {\mathcal{H}} _{B}\rightarrow {\mathcal{\ H}}_{C}$ is
a partial isometry. Denote $D_{C}= \sqrt{d_{B}}|S_{C}^{\ast }|$ and choose the
basis in which this operator is diagonal. Then (\ref{durep}) takes the form

\begin{equation*}
\tilde{\Phi}(\rho )=D_{C}W\left( \rho \otimes \frac{I_{B}}{d_{B}}\right)
W^{\ast }D_{C},
\end{equation*}
and (with some strain) can be interpreted as an interaction of the system $A$
with environment $B$ in the chaotic state followed by partial dephasing, cf.
\cite{DS}. Note however, that the \textquotedblleft
interaction\textquotedblright\ is only partially unitary and the dephasing
CP map is in general not TP (i.e. channel).

By interchanging the roles of ${\mathcal{H}}_{B},{\mathcal{H}}_{C}$ we of course
obtain a similar representation for the initial map $\Phi $
\begin{equation*}
\Phi (\rho )=S_{B}(\rho \otimes I_{C})S_{B}^{\ast },
\end{equation*}
where $S_{B}:{\mathcal{H}}_{A}\otimes {\mathcal{H}}_{C}\rightarrow
{\mathcal{H}}_{B}$ (in the case of channel $\mathrm{Tr}_{{\mathcal{H}}
_{C}}S_{B}^{\ast }S_{B}=I_{A}).$ This representation is especially nice in the case
where $A=B$ and $\Phi $ is unital: then $S_{B}$ is co-isometry, $ S_{B}S_{B}^{\ast
}=I_{A}.$

\bigskip The Kraus representation for the map $\Phi $
\begin{equation}
\Phi (\rho )=\sum_{j=1}^{d_{C}}V_{j}\rho V_{j}^{\ast };\qquad \rho \in
\mathfrak{\ \ M}(\mathcal{H}_{A}),  \label{kraus3}
\end{equation}
follows from (\ref{compl}) by letting $V_{j}=\langle e_{j}^{C}|V$ where $ \left\{
e_{j}^{C}\right\} $ is an orthonormal basis in $\mathcal{H}_{C}.$ Conversely,
\begin{equation*}
V=\sum_{j=1}^{d_{C}}V_{j}\otimes |e_{j}^{C}\rangle ,
\end{equation*}
whence, applying the second relation in (\ref{compl}), we have explicit
formula for the complementary map
\begin{equation}
\tilde{\Phi}(\rho )=\sum_{j,k=1}^{d_{C}}|e_{j}^{C}\rangle \langle e_{k}^{C}|
\mathrm{Tr}V_{j}\rho V_{k}^{\ast };\qquad \rho \in \mathfrak{M}(\mathcal{H} _{A}).
\label{cmp}
\end{equation}
It follows that the Kraus representation for $\tilde{\Phi}$ is
\begin{equation*}
\tilde{\Phi}(\rho )=\sum_{k=1}^{d_{B}}\tilde{V}_{k}\rho \tilde{V}_{k}^{\ast
},
\end{equation*}
where $\tilde{V}_{k}:\mathcal{H}_{A}\rightarrow \mathcal{H}_{C}$ are given
by
\begin{equation*}
\tilde{V}_{k}=\sum_{j=1}^{d_{C}}\langle e_{k}^{B}|V_{j}\otimes
|e_{j}^{C}\rangle ,
\end{equation*}
and hence satisfy
\begin{equation}
\langle e_{j}^{C}|\tilde{V}_{k}=\langle e_{k}^{B}|V_{j}.  \label{rec}
\end{equation}
The representation (\ref{durep}) takes place with
\begin{equation*}
S_{C}=\sum_{k=1}^{d_{B}}\tilde{V}_{k}\otimes \langle e_{k}^{B}|.
\end{equation*}

Finally, consider the case  where $A=B$ and $\Phi $ is unital, which is equivalent
to
\begin{equation*}
\sum_{j=1}^{d_{C}}V_{j}V_{j}^{\ast }=I_{A}.
\end{equation*}
By using (\ref{rec}) we obtain that this is the same as
\begin{equation*}
\mathrm{Tr}\tilde{V}_{j}^{\ast }\tilde{V}_{k}=\delta _{jk}.
\end{equation*}
Since $S_{C}^{\ast }S_{C}=\sum_{j,k=1}^{d_{B}}\tilde{V}_{j}^{\ast }\tilde{V}
_{k}\otimes |e_{j}^{B}\rangle \langle e_{k}^{B}|,$ this is equivalent to $
\mathrm{Tr}_{{\mathcal{H}}_{B}}S_{C}^{\ast }S_{C}=I_{A}.$

\section{Depolarizing channel}

Consider the depolarizing channel
\begin{equation}
\Phi (\rho )=(1-p)\rho +\frac{p}{d}I\mathrm{Tr}\rho ,\quad 0\leq p\leq \frac{
d^{2}}{d^{2}-1},
\end{equation}
where $\rho \in \mathfrak{M}\left( \mathcal{H}\right) $, with $\mathcal{H}
\simeq \mathbf{C}^{d}$. If $\{|j\rangle \,:j=1,\ldots ,d\}$ is a complete
set of orthonormal basis vectors in $\mathcal{H}$, then writing the channel
as
\begin{equation*}
\Phi (\rho )=(1-p)\rho +\frac{p}{d}\sum_{i,j=1}^{d}|i\rangle \langle j|\rho
| {j}\rangle \langle {i}|,
\end{equation*}
yields a Kraus representation with the operators
\begin{equation*}
V_{0}=\sqrt{1-p}I,\quad V_{ij}=\sqrt{\frac{p}{d}}|i\rangle \langle j|.
\end{equation*}

Let us relabel these Kraus operators as follows. Define a variable

\begin{equation*}
c(i,j):=i + (j-1)d,\quad 1\leq i,j\leq d
\end{equation*}
which takes integer values from $1$ to $d^{2}$. Then the Kraus operators can
be denoted as $A_{k}$, $0\leq k\leq d^{2}$, where
\begin{eqnarray}
A_{0}:= &&V_{0}\quad {\hbox{ and}}  \notag \\
A_{c(i,j)}:= &&V_{ji}\quad 1\leq i,j\leq d.  \label{krausdep}
\end{eqnarray}
Note that $A_{c(i,j)}^{\ast }=V_{ji}^{\ast }=V_{ij}$. The channel
complementary to the depolarizing channel is given by \cite{holcc}
\begin{equation*}
{\tilde{\Phi}}(\rho )=\Bigl[\tr \,A_{\alpha }\rho A_{\beta }^{\ast }\Bigr]
_{\alpha ,\beta =0,1,\ldots ,d^{2}}.
\end{equation*}
It is easy to see that
\begin{eqnarray}
\tr A_{0}\rho A_{0}^{\ast } &=&(1-p)\tr\,(I\rho I)=(1-p)\tr\rho ;  \notag \\
\tr A_{0}\rho A_{c(i,j)}^{\ast } &=&\sqrt{(1-p)}\tr(\rho V_{ij})=\sqrt{\frac{
p(1-p)}{d}}\langle j|\rho |i\rangle ;  \notag \\
\tr A_{c(i,j)}\rho A_{0}^{\ast } &=&\sqrt{\frac{p(1-p)}{d}}\langle i|\rho
|j\rangle ;  \notag \\
\tr A_{c(i,j)}\rho A_{c(i^{\prime },j^{\prime })}^{\ast } &=&\tr V_{ji}\rho
V_{i^{\prime }j^{\prime }}=\frac{p}{d}\langle i|\rho |i^{\prime
}\rangle\delta _{jj^{\prime }}  \label{relns1}
\end{eqnarray}

To express the complementary channel in a compact form, let us define a $
d^{2}$--dimensional row vector\footnote{ Here and henceforth, we use the notation
$|ij\rangle $ to denote the vector $ |i\rangle \otimes |j\rangle $. Consequently,
$\langle ij|=\langle i\otimes \langle j|$.}
\begin{equation}
\vec{\rho}:=\sum_{i,j=1}^{d^{2}}\rho _{ji}\langle ij|,\quad {\hbox{where }}
\quad \rho _{ji}=\langle j|\rho |i\rangle .
\end{equation}
In terms of this vector and its transpose $\vec{\rho}^{T}$, the complementary
channel $\tilde{\Phi}(\rho )$ can be represented by a $ (d^{2}+1)\times (d^{2}+1)$
matrix
\begin{equation}
\tilde{\Phi}(\rho )=\left[
\begin{array}{cc}
(1-p)\mathrm{Tr}\rho & \sqrt{\frac{p(1-p)}{d}}\vec{\rho} \\
\sqrt{\frac{p(1-p)}{d}}{\vec{\rho}}^{{\small {T}}} & \frac{p}{d}(\rho \otimes I)
\end{array}
\right] .  \label{matdep}
\end{equation}
This representation is not minimal since the number of Kraus operators $
A_{k} $ (defined by (\ref{krausdep})) is $d^{2}+1$. However, a minimal
representation for $\tilde{\Phi}$ can be obtained from (\ref{matdep}) as
follows. Note that (\ref{matdep}) can be equivalently written as
\begin{equation*}
\tilde{\Phi}(\rho )=T\rho T^{\ast },
\end{equation*}
where
\begin{equation*}
T^{\ast }=\left[
\begin{array}{cc}
\sqrt{d(1-p)}|\Omega _{12}\rangle & \sqrt{\frac{p}{d}}I_{12},
\end{array}
\right] .
\end{equation*}
with $|\Omega _{12}\rangle =d^{-1/2}\sum_{j=1}^{d}|{j}{j}\rangle $ the maximally
entangled vector in ${\mathcal{H}}\otimes {\mathcal{H}}$ and $ I_{12}$ is the
identity operator in ${\mathcal{H}}\otimes {\mathcal{H}}$. Let $T=US$ be its polar
decomposition, where $S=|T|=\sqrt{T^{\ast }T}$ is a positive Hermitian operator in
${\mathcal{H}}\otimes {\ \mathcal{H\simeq H}} _{d^{2}}$ and $U$ is an isometry from
${\mathcal{H}}_{d^{2}}$ to ${\mathcal{ H }}_{d^{2}+1},$ which is irrelevant for the
minimal representation we are looking for. Since
\begin{equation*}
T^{\ast }T=\frac{p}{d}I_{12}+d(1-p)|\Omega _{12}\rangle \langle \Omega _{12}|
\end{equation*}
is easily diagonalizable, we find
\begin{equation*}
S=\sqrt{T^{\ast }T}=\sqrt{\frac{p}{d}}I_{12}+\sqrt{d} \Bigl(-\frac{\sqrt{p}}{
d}+\sqrt{1-p\bigl(\frac{d^{2}-1}{d^{2}}\bigr)} \Bigr)|\Omega _{12}\rangle
\langle \Omega _{12}| ,
\end{equation*}
and the minimal representation of the complementary channel is
\begin{equation}  \label{dc}
\tilde{\Phi}(\rho )=S(\rho \otimes I )S^{\ast }
\end{equation}

While the depolarizing channel is globally unitarily covariant, the
complementary channel has the covariance property
\begin{equation*}
\tilde{\Phi}[U\rho U^{\ast }]=(U\otimes \bar{U})\tilde{\Phi}[\rho ](U\otimes
\bar{U})^{\ast }
\end{equation*}
for arbitrary unitary operator $U$ in $\mathcal{H}$.

By the results in \cite{holcc}, \cite{rus}, the complementary channel (\ref {dc})
has the same multiplicativity/additivity properties as the depolarizing channel
established in \cite{king}.

\section{Transpose-depolarizing channel}

Consider the one-parameter family of channels in ${\mathcal{H}} \simeq
\mathbf{C}^d$
\begin{equation}
\Phi(\rho) = t \rho^T + (1-t) \tr \rho \frac{{I}}{d},  \label{channel}
\end{equation}
where
\begin{equation}
- \frac{1}{d - 1} \le t \le \frac{1}{d+1}.  \label{range2}
\end{equation}
Here $\rho^{T}$ denotes transpose of the matrix $\rho$ in a fixed basis. The
channel $\Phi $ is irreducibly covariant since for any arbitrary unitary
transformation $U$
\begin{equation}
\Phi (U\rho U^{\ast })=\bar{U}\Phi (\rho )\bar{U}^{\ast },  \label{cov}
\end{equation}
where $\bar{U}$ is the complex conjugate of $U$ in the fixed basis. For this
class of channels, additivity of the minimum output entropy and the
multiplicativity of its maximal $p$--norm for $1 \le p \le 2$, has been
proved in \cite{fannes, dhstd, ndmult}. As it was shown in \cite{dhstd},
this channel can also be written as
\begin{equation}
\Phi(\rho) = c^+\Phi^+(\rho) + c^-\Phi^-(\rho),  \label{td}
\end{equation}
where
\begin{equation*}
c^{\pm}= \left(\frac{d^2 - 1}{2d}\right)\left( \frac{1}{d \mp 1} \pm
t\right),
\end{equation*}
and
\begin{equation}
\Phi^\pm (\rho):= \frac{1}{d\pm 1}\left({I} \tr \rho\pm \rho^T\right).
\label{whdef}
\end{equation}

Note that the extreme channel $\Phi^-(\rho)$ is the well known Werner-Holevo
(WH) channel \cite{WH}. The channels $\Phi^\pm (\rho)$ have Kraus operators
\begin{equation}
V^{\pm}_{ij} := \frac{1}{\sqrt{2(d \pm 1)}}\left(|i\rangle\langle j| \pm
|j\rangle\langle i|\right),
\end{equation}
where $|i\rangle, |j\rangle$ denote orthonormal basis vectors in ${\mathcal{ H }}$.
Let us relabel these operators using the variable
\begin{equation*}
c(i,j) = i+ (j-1)d , \quad 1 \le i \le d, \, 1 \le j \le 2d.
\end{equation*}
and the relations
\begin{equation*}
A^+_{c(i,j)} = \sqrt{c^+} V^+_{ji} \quad {\hbox{for }}\,\, 1 \le i,j, \le d;
\end{equation*}
\begin{equation*}
A^-_{c(i,j)} =\sqrt{c^-} V^-_{(j-d)\,i} \quad {\hbox{for }}\,\, 1 \le i \le
d, \, \, (d+1) \le j \le 2d.
\end{equation*}
Note that $c(i,j)$ takes integer values from $1$ to $2d^2$. In terms of the
above operators, the Kraus operators of the transpose depolarizing channel $
\Phi$, (\ref{td}), can be expressed as
\begin{equation}
A_{c(i,j)} := A^+_{c(i,j)} {\mathcal{I}}(1 \le j \le d) + A^-_{c(i,j)}{\
\mathcal{I}}(d+1 \le j \le 2d),  \label{kraustd}
\end{equation}
where ${\mathcal{I}}(\cdot )$ denotes an indicator function. Its
complementary channel is given by
\begin{equation*}
\tilde{\Phi}(\rho ) := \Bigl[\tr A_\alpha \rho A_\beta^* \Bigr]_{\alpha,
\beta = 1, \ldots, 2d^2}.
\end{equation*}

Let us first consider the case $1\leq \alpha ,\beta \leq d^{2}$, for which $ \alpha
=c(i,j)$ and $\beta =c(i^{\prime },j^{\prime })$ for some $1\leq i,i^{\prime }\leq
d$ and $1\leq j,j^{\prime }\leq d$. From (\ref{kraustd}) it follows that
\begin{eqnarray*}
\tr A_{c(i,j)}\rho A_{c(i^{\prime },j^{\prime })}^{\ast } &=& \tr \,{c^{+}}
V_{ji}^{+}\rho V_{i^{\prime }j^{\prime }}^{+} \\
&=& \frac{c^{+}}{2(d+1)} \left[ \delta _{jj^{\prime }}\rho _{ii^{\prime
}}+\delta _{ji^{\prime }}\rho _{ij^{\prime }}+\delta _{ij^{\prime }}\rho
_{ji^{\prime }}+\delta _{ii^{\prime }}\rho _{jj^{\prime }}\right] . \\
&=& \frac{c^{+}}{2(d+1)} \left[ \rho\otimes I +(\rho\otimes I
)F+F(\rho\otimes I )+F(\rho \otimes I )F\right] _{ij,i^{\prime }j^{\prime }}
\\
&=& \frac{c^{+}}{2(d+1)} \left[ (I_{12}+F)(\rho \otimes I )(I_{12}+F)\right]
_{ij,i^{\prime }j^{\prime }}
\end{eqnarray*}
Here the flip operator $F$ is defined by its action
\begin{equation*}
F|ij\rangle =|ji\rangle ,
\end{equation*}
on basis vectors $|ij\rangle $ in ${\mathcal{H}}\otimes {\mathcal{H}}$ and $
I_{12}$ is the identity operator in ${\mathcal{H}}\otimes {\mathcal{H}}$.
Moreover, $\rho _{ij}:=\langle i|\rho |j\rangle $.

Similarly, for $(d^{2}+1)\leq \alpha ,\beta \leq 2d^{2}$ we have $\alpha =c( i,
\tilde{\jmath})$ and $\beta =c(i^{\prime },{\tilde{\jmath}})$ for some $ 1\leq
i,i^{\prime }\leq d$ and $d+1\leq \tilde{\jmath},{\tilde{\jmath}} ^{\prime }\leq
2d$. Defining $j=\tilde{\jmath}-d$ and $j^{\prime }={\ \tilde{ \jmath}}^{\prime
}-d$, we get
\begin{eqnarray*}
\tr A_{c(i, {\tilde{\jmath}})}\rho A_{c(i^{\prime}, {\tilde{\jmath}}^{\prime
})} ^{\ast } &=& \tr\,{c^{-}}V_{ji}^{-}\rho V_{i^{\prime }j^{\prime }}^{-} \\
&=& \frac{c^{-}}{2(d-1)} \left[ \delta _{jj^{\prime }}\rho _{ii^{\prime
}}-\delta _{ji^{\prime }}\rho _{ij^{\prime }}-\delta _{ij^{\prime }}\rho
_{ji^{\prime }}+\delta _{ii^{\prime }}\rho _{jj^{\prime }}\right] \\
&=&\frac{c^{-}}{2(d-1)}\left[ (I_{12}-F)(\rho \otimes I )(I_{12}-F)\right]
_{ij,i^{\prime }j^{\prime }}.
\end{eqnarray*}

For $\alpha =c({i},j)$, $\beta =c(i^{\prime }, {\tilde{\jmath}}^{\prime })$ for
some $1\leq i,i^{\prime },j^{\prime }\leq d$ and $d+1\leq {\tilde{\jmath} }^{\prime
}\leq 2d$, we have
\begin{eqnarray*}
\tr A_{c({i},j)}\rho A_{c(i^{\prime }, {\tilde{\jmath}}^{\prime })}^{\ast }
&=&\tr \,\sqrt{c^{+}c^{-}}V_{ji}^{+}\rho V_{i^{\prime }j^{\prime }}^{-} \\
&=&\frac{1}{2}\sqrt{\frac{c^{+}c^{-}}{(d+1)(d-1)}}\left[ (I_{12}+F)(\rho
\otimes I )(I_{12}-F)\right] _{ij,i^{\prime }j^{\prime }}.
\end{eqnarray*}

By symmetry, for $\alpha =c(i, \tilde{\jmath})$, $\beta =c({i}^{\prime },j^{\prime
})$ for some $1\leq i, i^{\prime },j^{\prime }\leq d$ and $ d+1\leq
\tilde{\jmath}\leq 2d$, we have
\begin{equation*}
\tr A_{c(i, {\tilde{\jmath}})}\rho A_{c({i}^{\prime },j^{\prime })}^{\ast }=
\frac{1}{2}\sqrt{\frac{c^{+}c^{-}}{(d+1)(d-1)}}\left[ (I_{12}-F)(\rho
\otimes I )(I_{12}+F)\right] _{ij,i^{\prime }j^{\prime }}.
\end{equation*}

From the above relations one concludes that the complementary channel of the
transpose-depolarizing channel has the (non--minimal) representation
\begin{equation*}
{\tilde{\Phi}}(\rho )=T(\rho \otimes I )T^{\ast },
\end{equation*}
where
\begin{equation*}
T^{\ast }=\left[
\begin{array}{cc}
a^{+}(I_{12}+F) & a^{-}(I_{12}-F)
\end{array}
\right] .
\end{equation*}
with
\begin{equation*}
a^{\pm }:=\sqrt{\frac{c^{\pm }}{2(d\pm 1)}}.
\end{equation*}

Let $T=US$ denote the polar decomposition of the matrix $T$, where $S=|T|=
\sqrt{T^{\ast }T}$ is a positive Hermitian operator in ${\mathcal{H}}\otimes
{\ \mathcal{H\simeq H}}_{d^{2}}$ and $U$ is an isometry from ${\mathcal{H}}
_{d^{2}}$ to ${\mathcal{H}}_{2d^{2}}$. By using the fact that $(I_{12}\pm
F)/2$ are projection operators we obtain the minimal representation
\begin{equation}  \label{ctd}
{\tilde{\Phi}}(\rho )=S(\rho \otimes I )S^{\ast },
\end{equation}
where
\begin{equation*}
S=\sqrt{T^{\ast }T}=(a^{+}+a^{-})I_{12}+(a^{+}-a^{-})F.
\end{equation*}

The covariance property of the channel (\ref{ctd}) is
\begin{equation*}
\tilde{\Phi}(U\rho U^{\ast })=(U\otimes U)\tilde{\Phi}(\rho )(U^{\ast
}\otimes U^{\ast }),
\end{equation*}
as follows from the fact that $F(U\otimes U)=(U\otimes U)F.$

\section{\protect\bigskip Coupling channel with its complementary}

Let us now study the properties of a channel which is a tensor product of
the WH channel
\begin{equation}
\Phi (\rho ):=\frac{1}{d-1}\left( {I}\tr\rho -\rho ^{T}\right) ,
\end{equation}
and the complementary channel
\begin{equation}
{\tilde{\Phi}}(\rho )=\frac{1}{2(d-1)}(I_{12}-F)(\rho \otimes I )(I_{12}-F).
\label{whcomp}
\end{equation}
The particular significance of the WH channel lies in the fact that it provides a
counterexample for the multiplicativity of the maximal output $p$ -norm for
$p>4.79$ and $d=3$ \cite{WH}. It is interesting to investigate whether a similar
violation of multiplicativity is exhibited for the product channel $\Phi \otimes
{\tilde{\Phi}}$. The multiplicativity of the maximal output $p$-norm and hence,
additivity of the minimum output R\'{e}nyi $p-$ entropies of the WH channel for
$p\in \lbrack 1,2]$ was established in \cite {my, af, dhs}. It is also interesting
to study whether these additivity properties hold for the channel $\Phi \otimes
\tilde{\Phi}$.

For the WH channel, $\check{R}_{p}(\Phi )=\check{R}_{p}(\tilde{\Phi})=\log (d-1)$
for all $p\geq 1$, since $\nu_p(\Phi)=(d-1)^{(1-p)/p}$ as shown in \cite{WH}.
Further, it was observed in \cite{WE} that if for some channel $ \Phi $
\begin{equation*}
\check{R}_{p}(\Phi )=\check{R}_{q}(\Phi )\quad {\hbox{ for }}1\leq q\leq p,
\end{equation*}
then the additivity of the minimal output R\'{e}nyi $p$-entropy implies the
additivity of the minimal output R\'enyi $q$-entropy. By using these facts,
the proof of the additivity relation
\begin{equation}
\check{R}_{p}(\Phi \otimes {\tilde{\Phi}})=\check{R}_{p}(\Phi )+\check{R}
_{p}(\tilde{\Phi})  \label{addprod}
\end{equation}
reduces to proving
\begin{equation}
\check{R}_{2}(\Phi \otimes {\tilde{\Phi}})=2\check{R}_{2}(\Phi ).
\label{addren}
\end{equation}
We can restate the additivity conjecture (\ref{addren}) as a
multiplicativity of maximal $2$--norms
\begin{equation}
\nu _{2}(\Phi \otimes {\tilde{\Phi}})=\nu _{2}(\Phi )\nu _{2}({\tilde{\Phi}}
)=\nu _{2}(\Phi )^{2},  \label{final}
\end{equation}
where
\begin{equation}
\nu_2 (\Phi \otimes {\tilde{\Phi}}) := \max_{\atop{|\psi_{12}\rangle \in {\
\mathcal{H}}_1 \otimes {\mathcal{H}}_2 }{{\ ||\psi_{12} ||=1}}}\, \left\{
||(\Phi\otimes {\tilde{\Phi}})(|\psi_{12}\rangle\langle \psi_{12}| )||_p\right\},
\label{multten}
\end{equation}
and we have made use of the relation $\nu _{2}({\tilde{\Phi}})=\nu _{2}(\Phi
)$ \cite{holcc}. To prove (\ref{final}), it is sufficient to show that the
maximum on the right hand side of (\ref{multten}) is achieved for
unentangled vectors $|\psi _{12}\rangle $, which in turn corresponds to the
reduced states $\rho _{1}:=\tr_{{\mathcal{H}}_{2}}|\psi _{12}\rangle \langle
\psi _{12}|$ and $\rho _{2}:=\tr_{{\mathcal{H}}_{1}}|\psi _{12}\rangle
\langle \psi _{12}|$ being pure.

Let the output of the product channel for an arbitrary pure input state $
|\psi _{12}\rangle \langle \psi _{12}|\in \mathfrak{S}({\mathcal{H}}
_{1}\otimes {\mathcal{H}}_{2})$, be denoted by
\begin{equation}
\Omega :=(\Phi \otimes {\tilde{\Phi}})(|\psi _{12}\rangle \langle \psi
_{12}|)=(\mathrm{Id}\otimes {\tilde{\Phi}})(\Phi \otimes \mathrm{Id})(|\psi
_{12}\rangle \langle \psi _{12}|),
\end{equation}
where $\mathrm{Id}$ is the identity channel. Due to the unitary covariance
of the channel $\Phi \otimes {\tilde{\Phi}}$, the state vector $|\psi
_{12}\rangle $ can be chosen as
\begin{equation}
|\psi _{12}\rangle =\sum_{j=1}^{d}\sqrt{\lambda _{j}}|j\rangle \otimes
|j\rangle ,  \label{schmidt}
\end{equation}
where $\{|j\rangle \}$ is the fixed orthonormal basis in $\mathbf{C}^{d}$
(one which defines the transposition), $\lambda _{j}\geq 0$ and $
\sum_{j=1}^{d}\lambda _{j}=1$. The reduced density matrices $\rho
_{i},\,i=1,2$ are therefore given by
\begin{equation}
\rho :=\rho _{1}=\sum_{j=1}^{d}\lambda _{j}|j\rangle \langle j|=\rho _{2}.
\label{red}
\end{equation}

Using the decomposition (\ref{schmidt}) we find that
\begin{equation*}
(\Phi \otimes \mathrm{Id})(|\psi _{12}\rangle \langle \psi _{12}|)=
\sum_{j,k} \sqrt{\lambda_j \, \lambda_k} \Phi(|j\rangle \langle k|) \otimes
|j\rangle \langle k|.
\end{equation*}
From the definition (\ref{whdef}) of the WH channel it follows that
\begin{equation*}
\Phi(|j\rangle \langle k|) = \frac{1}{d-1}\left(I \delta_{jk} - |k\rangle
\langle j| \right),
\end{equation*}
which in turn implies that
\begin{eqnarray*}
(\Phi \otimes \mathrm{Id})(|\psi _{12}\rangle \langle \psi _{12}|)&=& \frac{1
}{d-1}\left[\sum_j \lambda_j I \otimes |j\rangle \langle j| - \sum_{j,k}
\sqrt{\lambda_j \lambda_k} |kj\rangle \langle jk \right]  \notag \\
&=& \frac{1}{d-1}\left[I_{12}\otimes \rho - F(\sqrt{\rho} \otimes \sqrt{\rho}
)\right],
\end{eqnarray*}
where $\rho$ is given by (\ref{red}) and hence $\sqrt{\rho} =\sum_j \sqrt{
\lambda_j} |j\rangle \langle j|$.

Due to the relation $F(I\otimes \rho )=(\rho \otimes I)F$, the complementary
channel (\ref{whcomp}) can be alternatively expressed in the following
forms:
\begin{eqnarray*}
{\tilde{\Phi}}(\rho )&=&\frac{1}{(d-1)}\left( \frac{I_{12}-F}{2}\right)
\left( \rho \otimes I+I \otimes \rho\right)  \notag \\
&=& \frac{1}{(d-1)}\left( \frac{I_{12}-F}{2}\right) \left( \rho \otimes
I+F(\rho \otimes I)F\right)  \label{wh2}
\end{eqnarray*}
Using the above relations we get
\begin{eqnarray}
\Omega &=&(\mathrm{Id}\otimes {\tilde{\Phi}})\left[ \frac{1}{d-1}\Bigl(
I_{12}\otimes \rho -F_{12}( \sqrt{\rho }\otimes \sqrt{\rho })\Bigr) \right]
\notag \\
&=&\frac{1}{(d-1)^{2}}\left( \frac{I_{123}-F_{23}}{2}\right) \Bigl[I\otimes
\rho \otimes I+I\otimes I\otimes \rho  \notag \\
&-&F_{12}(\sqrt{\rho }\otimes \sqrt{\rho }\otimes I)-F_{23}(F_{12}(\sqrt{
\rho }\otimes \sqrt{\rho }\otimes I))F_{23}\Bigr],  \label{step2}
\end{eqnarray}
where we have defined
\begin{equation*}
I_{123}=I_{12}\otimes I,\quad F_{23}:=I\otimes F,\quad F_{12}:=F\otimes I.
\end{equation*}
we can now evaluate $\tr \Omega^2$ by employing the spectral decompositions of
$\rho$ (and hence of $\sqrt{\rho}$), the resolution of the identity $I = \sum_k
|k\rangle \langle k|$, and the explicit actions of the operators $ F_{12}$ and
$F_{23}$ on basis vectors, namely,
\begin{equation*}
F_{12}|ijk\rangle = |jik\rangle; \quad F_{23}|ijk\rangle = |ikj\rangle,
\end{equation*}
where $|ijk\rangle :=|i\rangle \otimes |j\rangle \otimes |k\rangle $. This
calculation yields
\begin{equation*}
\tr \Omega^2 = \frac{1}{(d-1)^4} \left[ (d^2 - 4d + 5)\tr \rho^2 + 2(d-2) \right],
\end{equation*}
which is indeed maximised when $\tr \rho^2 = 1$, i.e., when $\rho$ is a pure
state. Thus we see that, for the product channel $\Phi \otimes {\tilde{\Phi}}
$, the multiplicativity (\ref{final}) of the $2$--norms and hence the
additivity (\ref{addprod}) of the minimum output entropy holds.

To investigate violation of multiplicativity for the product channel $\Phi
\otimes {\tilde{\Phi},}$ let us consider the output $\Omega ^{me}$ of this
channel when the input is the maximally entangled state $|\psi _{me}\rangle
\langle \psi _{me}|$,
\begin{equation*}
|\psi _{me}\rangle =\frac{1}{\sqrt{d}}\sum_{j=1}^{d}|jj\rangle .
\end{equation*}
In this case the reduced density matrix $\rho $, defined by (\ref{red}), is
the completely mixed state: $\rho =I/d$. Hence, $\Omega ^{me}$ is simply
obtained from (\ref{step2}) by replacing $\rho $ by $I/d$ on its right hand
side. This yields the relation
\begin{eqnarray*}
\Omega ^{me} &=&\frac{1}{d(d-1)^{2}}\Bigl[P_{1} \\
&+&\frac{1}{2}\bigl(
I_{123}-F_{23}-F_{12}-F_{23}F_{12}F_{23}+F_{23}F_{12}+F_{12}F_{23}\bigr) \Bigr ].
\end{eqnarray*}
where $P_{1}:=(I_{123}-F_{23})/{2}$ is a projection operator.

Let us express $\Omega ^{me}$ in a more transparent form, in order to
evaluate its eigenvalues. For this purpose, define a vector
\begin{equation*}
|\phi _{\{ijk\}}\rangle :=\frac{1}{\sqrt{6}}\Bigl[|ijk\rangle +|jki\rangle
+|kij\rangle -|jik\rangle -|kji\rangle -|ikj\rangle \Bigr].
\end{equation*}
It is of unit norm and satisfies the relations
\begin{eqnarray*}
F_{12}|\phi _{\{ijk\}}\rangle &=&-|\phi _{\{ijk\}}\rangle \\
F_{23}|\phi _{\{ijk\}}\rangle &=&-|\phi _{\{ijk\}}\rangle
\end{eqnarray*}
Moreover,
\begin{equation*}
\langle \phi _{\{ijk\}}|\phi _{\{i^{\prime }j^{\prime }k^{\prime }\}}\rangle
=0\quad {\hbox{unless}}\quad \{ijk\}=\{i^{\prime }j^{\prime }k^{\prime }\},
\end{equation*}
and hence the set of vectors
\begin{equation*}
\Bigl\{|\phi _{\{ijk\}}\rangle \,:\,i,j,k\in \{1,2,\ldots ,d\},i,j,k\,\,{\
\hbox{all
different}}\Bigr\}
\end{equation*}
form an orthonormal set. Therefore
\begin{equation*}
P_{2}:= \sum_{\atop{\{ijk\}}{{\atop{i,j,k \in \{1,2, \ldots, d\}}{{\ i,j,k \,\,
{\hbox{all different}}}}}}} |\phi_{\{ijk\}}\rangle\langle \phi_{\{ijk\}}|,
\end{equation*}
is a projection operator. Moreover
\begin{equation*}
{\hbox{ran }}P_{2}\subset {\hbox{ran
}} P_{1}.
\end{equation*}
It is easy to see that
\begin{equation*}
I_{123}-F_{23}-F_{12}-F_{23}F_{12}F_{23}+F_{23}F_{12}+F_{12}F_{23}= 6P_{2}.
\end{equation*}
Hence,
\begin{equation*}
\Omega ^{me}=\frac{1}{d(d-1)^{2}}\left[ P_{1}+3P_{2}\right] ,
\end{equation*}
and its eigenvalues are

\begin{enumerate}
\item {${4}/{d(d-1)^2}$ with multiplicity
\begin{equation*}
{\binom{d }{3}} \equiv {\hbox{number of distinct subsets }}\, \{ijk\}\, {\
\hbox{of the
set}} \{1,2, \ldots, d\};
\end{equation*}
}

\item {\ ${1}/{d(d-1)^{2}}$ with multiplicity
\begin{equation*}
{\hbox{dim }}({\hbox{range of }}P_{1}\setminus P_{2})=\frac{d^{2}(d-1)}{2}-{ \
\binom{d}{3}}=\frac{d(d^{2}-1)}{3}
\end{equation*}
}

\item {\ $0$ with multiplicity $d(d+1)/2$ }
\end{enumerate}

For $d=3$, therefore, there is a non-degenerate eigenvalue of $1/3$, the eigenvalue
$1/12$ with multiplicity $8$, and the eigenvalue $0$ with multiplicity $6$. The
non--zero eigenvalues are found to be exactly identical those of the channel $\Phi
\otimes \Phi $ for $d=3$, (see \cite{WH} ), for which a violation of the
multiplicativity of the maximal output $p$ --norm was obtained for $p>4.79$. Hence,
we deduce that a similar violation of multiplicativity is exhibited for the channel
$\Phi \otimes {\tilde{\Phi}} $ for $p>4.79$ and $d=3$.

For $d\geq 4$ we get
\begin{equation*}
\nu _{p}(\Phi \otimes {\tilde{\Phi}})^{p}/\nu _{p}(\Phi )^{p}\nu _{p}({\
\tilde{\Phi}})^{p}\ge \frac{1}{(d-1)^{2}}\left[ {\binom{d}{3}}\left( {{4}/{d} }
\right) ^{p}+\frac{d(d^{2}-1)}{3}\left( {{1}/{d}}\right)^{p}\right],
\end{equation*}
but the right hand side is always less or equal than $1$ for $p\ge 1$, so contrary
to the case of $\Phi \otimes \Phi ,$ considering the output for the maximally
entangled state does not allow us to conclude violation of multiplicativity.
However, this might be due to the fact that the channel $ \Phi \otimes
{\tilde{\Phi}}$ does not have the flip symmetry of the channel $ \Phi \otimes \Phi
,$ and the maximizing input state could be different from the maximally entangled
one.

\textbf{Acknowledgments.} This work was accomplished when A. H. was the
Leverhulme Visiting Professor at DAMTP, CMS, University of Cambridge. The
authors are grateful to Yu. M. Suhov for useful discussions.

\end{document}